\documentclass{iaus}
\usepackage{graphicx}
  \checkfont{eurm10}
  \iffontfound
    \IfFileExists{upmath.sty}
      {\typeout{^^JFound AMS Euler Roman fonts on the system,
                   using the 'upmath' package.^^J}%
       \usepackage{upmath}}
      {\typeout{^^JFound AMS Euler Roman fonts on the system, but you
                   dont seem to have the}%
       \typeout{'upmath' package installed. iaus.cls can take advantage
                 of these fonts,^^Jif you use 'upmath' package.^^J}%
      }
  \else
  \fi

  \checkfont{msam10}
  \iffontfound
    \IfFileExists{amssymb.sty}
      {\typeout{^^JFound AMS Symbol fonts on the system, using the
                'amssymb' package.^^J}%
       \usepackage{amssymb}%

      }{}
  \fi

  \IfFileExists{amsbsy.sty}
    {\typeout{^^JFound the 'amsbsy' package on the system, using it.^^J}%
     \usepackage{amsbsy}}
    {}

\newsavebox{\astrutbox}
\sbox{\astrutbox}{\rule[-5pt]{0pt}{20pt}}

\newcommand\etal{\mbox{\textit{et al.}}}

\title[Outskirts of Galaxy Clusters: intense life in the suburbs]
      {Galaxy evolution in dense environments;\break a concise HI perspective.}

\author[M.A.W. Verheijen {\it et al.\/}]%
{Marc A.W. Verheijen$^1$%
}

\affiliation{$^1$Kapteyn Astronomical Institute, Postbus 800, 9700 AV
Groningen, The Netherlands \break email: verheyen@astro.rug.nl}

\pubyear{2004}
\volume{195}
\pagerange{1--8}
\date{?? and in revised form ??}
\jname{Outskirts of Galaxy Clusters: intense life in the suburbs}
\editors{A. Diaferio, ed.}
\begin{document}

\maketitle

\begin{abstract}
Observing the neutral Hydrogen in galaxy clusters provides crucial
insights in the physical processes that influence the evolution of
gas-rich galaxies as they migrate from the lower-density filaments
through the cluster outskirts into to the higher-density central
regions. The morphology-density relation, the Buther-Oemler effect,
and the observed HI deficiencies in the central regions of galaxy
clusters suggest that infalling galaxies experience a strong
transformation of their morphologies, star formation rates, and gas
content. The physical mechanisms that trigger and govern these
transformations may depend strongly on environment. This contribution
aims to illustrate that the morphological and kinematic
characteristics of the cold gas provide a sensitive tool to determine
which mechanisms dominate in which environments.
\end{abstract}

\firstsection % if your document starts with a section,
              % remove some space above using this command.

\section{Motivation for observing the cold gas.}\label{sec:sec1}

Hydrogen is the most abundant primordial baryonic matter in the
Universe out of which the visible galaxies and galaxy clusters have
formed. Observations of the distribution and kinematics of cold
neutral Hydrogen at the epoch of galaxy formation, and during their
subsequent evolution, provide one of the most direct windows on the
physical processes involved. Numerous studies indicate that the pace
of a galaxy's evolution, like its growth by mass accretion, the rate
at which it forms stars, the frequency of tidal interactions and the
occurrence of minor or major mergers, depends significantly on its
local and global environment. In particular the environment
encountered in the outskirts of galaxy clusters is very favorable to
rapid evolution and transformation of galaxies as they fall in along
the lower-density filaments into the higher-density cluster
cores. Such infalling galaxies encounter an increased local galaxy
density, which results in more frequent tidal interactions, and they
may be confronted with a hot Intra-Cluster Medium (ICM) as they
approach the cluster core. This change of environment can have a
dramatic impact on a galaxy's morphology, its gas content and its star
formation rate. This is also implied by the well-known
morphology-density relation combined with the notion that clusters
continue to accrete galaxies from the surrounding large scale
structure in which they are embedded. Consequently, infalling galaxies
must experience a signifiant transformation on a relatively short time
scale, which is also suggested by the Butcher-Oemler effect and the
interpretation that the blue galaxies in the outskirts of galaxy
clusters experience a period of vigorous star formation which can end
abruptly once all the cold gas is consumed or removed. These galaxies
eventually find themselves in the cluster core as gas-poor systems
with a passively evolving stellar population.

The presence of cold gas in galaxies is a prerequisite for star
formation, and the amount and distribution of cold gas regulates the
duration and intensity of star formation periods. The removal of this
cold gas reservoir from infalling galaxies is a crucial element in the
process that transforms these systems. The key question is which
physical processes trigger and govern the enhanced star formation, and
the removal of the cold gas. What is the role and ultimate fate of the
cold neutral Hydrogen in infalling galaxies?

\begin{table}
\caption{Galaxy clusters imaged in neutral Hydrogen, compiled from the literature and from work by the author and his collaborators.}
\begin{center}
\begin{tabular}{l|crrcrccc|cccc}

Cluster                      &
 z                           &
\multicolumn{1}{c}{$\sigma$} &
\multicolumn{1}{c}{N$_z$}    &
 T$_{\mbox{\tiny B$-$M}}$    &
\multicolumn{1}{c}{C}        &
 R                           &
 L$_X$             &
 f$_b$                       &
 Array                       &
 Mode                        &
 \# Flds                     &
 N$_{det}$                   \\
                &
                &
{\scriptsize km/s} & 
                &
                &
                &
                &
{\scriptsize 10$^{44}$erg/s}    &
                &
                &
                &
                &
                \\

\multicolumn{1}{c}{\scriptsize(1)}  &
\multicolumn{1}{c}{\scriptsize(2)}  &
\multicolumn{1}{c}{\scriptsize(3)}  &
\multicolumn{1}{c}{\scriptsize(4)}  &
\multicolumn{1}{c}{\scriptsize(5)}  &
\multicolumn{1}{c}{\scriptsize(6)}  &
\multicolumn{1}{c}{\scriptsize(7)}  &
\multicolumn{1}{c}{\scriptsize(8)}  &
\multicolumn{1}{c}{\scriptsize(9)}  &
\multicolumn{1}{c}{\scriptsize(10)} &
\multicolumn{1}{c}{\scriptsize(11)} &
\multicolumn{1}{c}{\scriptsize(12)} &
\multicolumn{1}{c}{\scriptsize(13)} \\

\hline
Ursa Major       & 0.003 &  149 &         96 &         &  26 &   &       &      & WSRT & pointed   & 65     & 70  \\
                 &       &      &            &         &     &   &       &      & VLA  & blind     & 56     &     \\
Virgo \hfill     & 0.004 &      &            &         &     &   &  1.00 & 0.04 & VLA  & pointed   & 23     & 23 \\
Hydra \hfill     & 0.013 &  676 &        256 & III     &  50 & 1 &  0.47 &      & VLA  & vol. lim. & 13     & 49  \\
Abell\hfill 3627 & 0.014 &  897 &        112 & I       &  59 & 1 &  2.2  &      & ATCA & pointed   &  3     &  2  \\
Abell\hfill  262 & 0.016 &  415 &      $>$38 & III     &  40 & 0 &  0.55 & 0.02 & WSRT & pointed   &  5     & 11  \\
Coma  \hfill     & 0.023 &  880 &$\sim$10$^3$& II      & 106 & 2 &  7.21 & 0.03 & VLA  & pointed   & 12     & 39  \\
                 &       &      &            &         &     &   &       &      & WSRT & blind     & mosaic &     \\
Abell\hfill  496 & 0.033 &  657 &        238 & I:      &  50 & 1 &  3.54 &      & VLA  & vol. lim. & 10     & 46  \\
Hercules \hfill  & 0.037 &  887 &      $>$42 & III     &  87 & 2 &  0.88 & 0.14 & VLA  & blind     &  4     & 61  \\
Abell\hfill  754 & 0.054 & 1048 &        408 & I-II:   &  92 & 2 &  8.01 &      & VLA  & vol. lim. &  2     &  4  \\
Abell\hfill   85 & 0.055 & 1443 &        275 & I       &  59 & 1 &  8.38 &      & VLA  & vol. lim. &  8     & 23  \\
Abell\hfill 2670 & 0.076 &  908 &        219 & I-II    & 142 & 3 &  2.55 & 0.04 & VLA  & vol. lim. &  3     & 49  \\
Abell\hfill 2218 & 0.176 &      &            & II:     & 214 & 4 &  8.99 & 0.11 & WSRT & vol. lim. &  1     &  1  \\
Abell\hfill 2192 & 0.188 &  706 &         36 & II-III: &  62 & 1 &       &      & VLA  & vol. lim. &  1     &  1  \\
                 &       &      &            &         &     &   &       &      & GMRT & vol. lim. &  1     &  0  \\
                 &       &      &            &         &     &   &       &      &      &           &        &     \\

Abell\hfill 1689 & 0.183 & 1253 &        525 & II-III: & 228 & 4 & 20.74 & 0.09 & VLA  & vol. lim. &  1     &  0  \\
Abell\hfill  963 & 0.206 &      &  $\sim$200 & I-II    & 134 & 3 & 10.23 & 0.19 & WSRT & vol. lim. &  1     &  0  \\
\hline

\multicolumn{13}{l}{\scriptsize Ursa Major: \cite{VS01}, \cite{VTTZ01}. Virgo: \cite{CvGBK90}.} \\
\multicolumn{13}{l}{\scriptsize Hydra: \cite{McM93}. A3627: \cite{VCvDHKBWD01}. A262: \cite{BSCBS97}.} \\
\multicolumn{13}{l}{\scriptsize Coma: \cite{BCvGB00}, \cite{B03}. A496: van Gorkom \etal\ (in progress).} \\
\multicolumn{13}{l}{\scriptsize Hercules: \cite{D97}. A754: van Gorkom \etal\ (in progress). A85: van Gorkom \etal\ (in progress).} \\
\multicolumn{13}{l}{\scriptsize A2670: \cite{PvG01}. A2218: \cite{ZvDV01}. A2192: Verheijen \etal\ (in progress).} \\
\multicolumn{13}{l}{\scriptsize A1689: Verheijen \etal\ (in progress). A963: van Gorkom \etal\ (in progress).} \\

\end{tabular}
\end{center}
\end{table}

\section{Physical processes.}\label{sec:sec2}

\begin{figure}

\begin{minipage}{6.5cm}
\includegraphics[width=6.5cm]{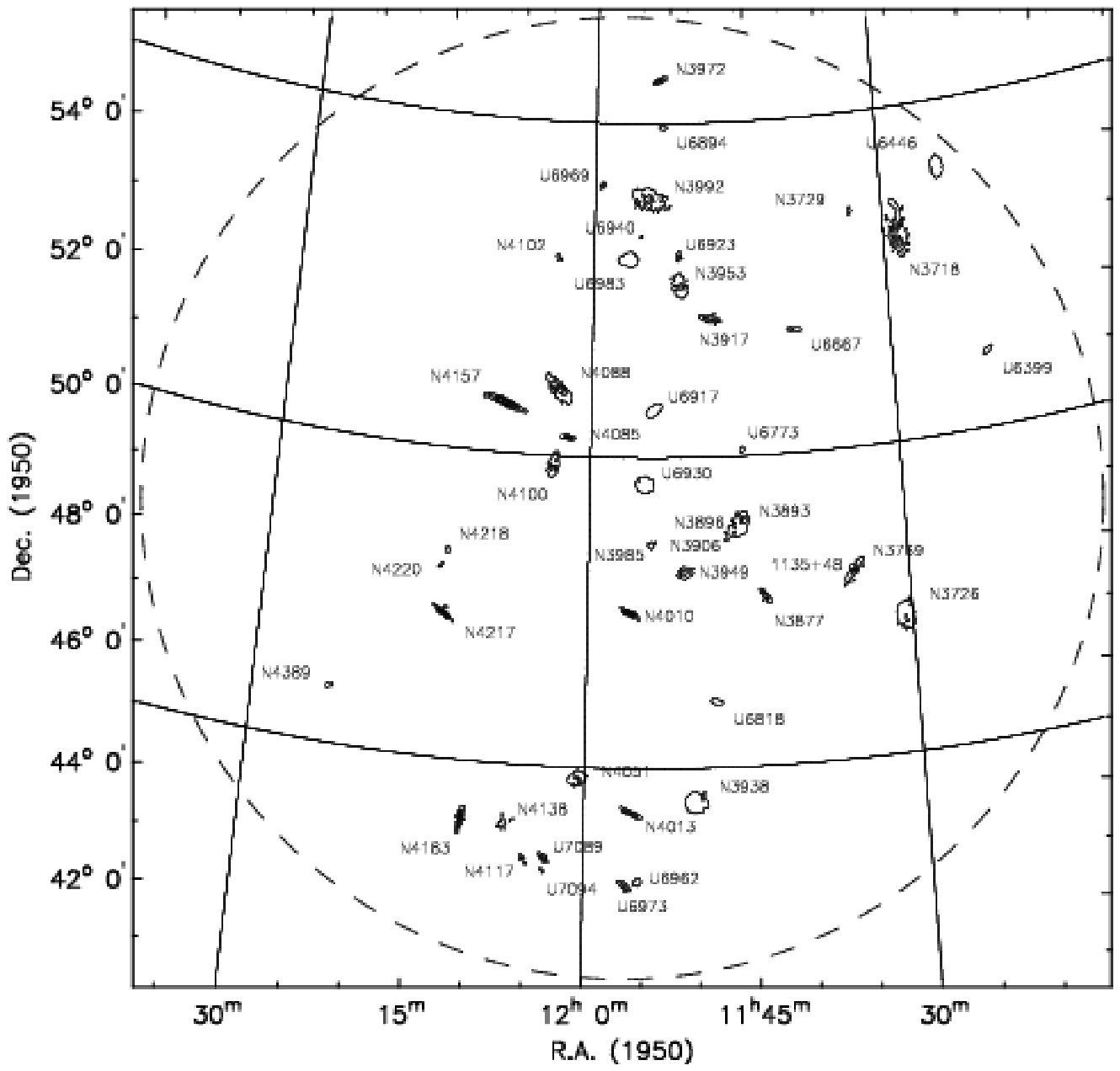}
\end{minipage}

\hspace{7.0cm}
\begin{minipage}{6.5cm}
\vspace{-6.2cm}
\includegraphics[width=6.5cm]{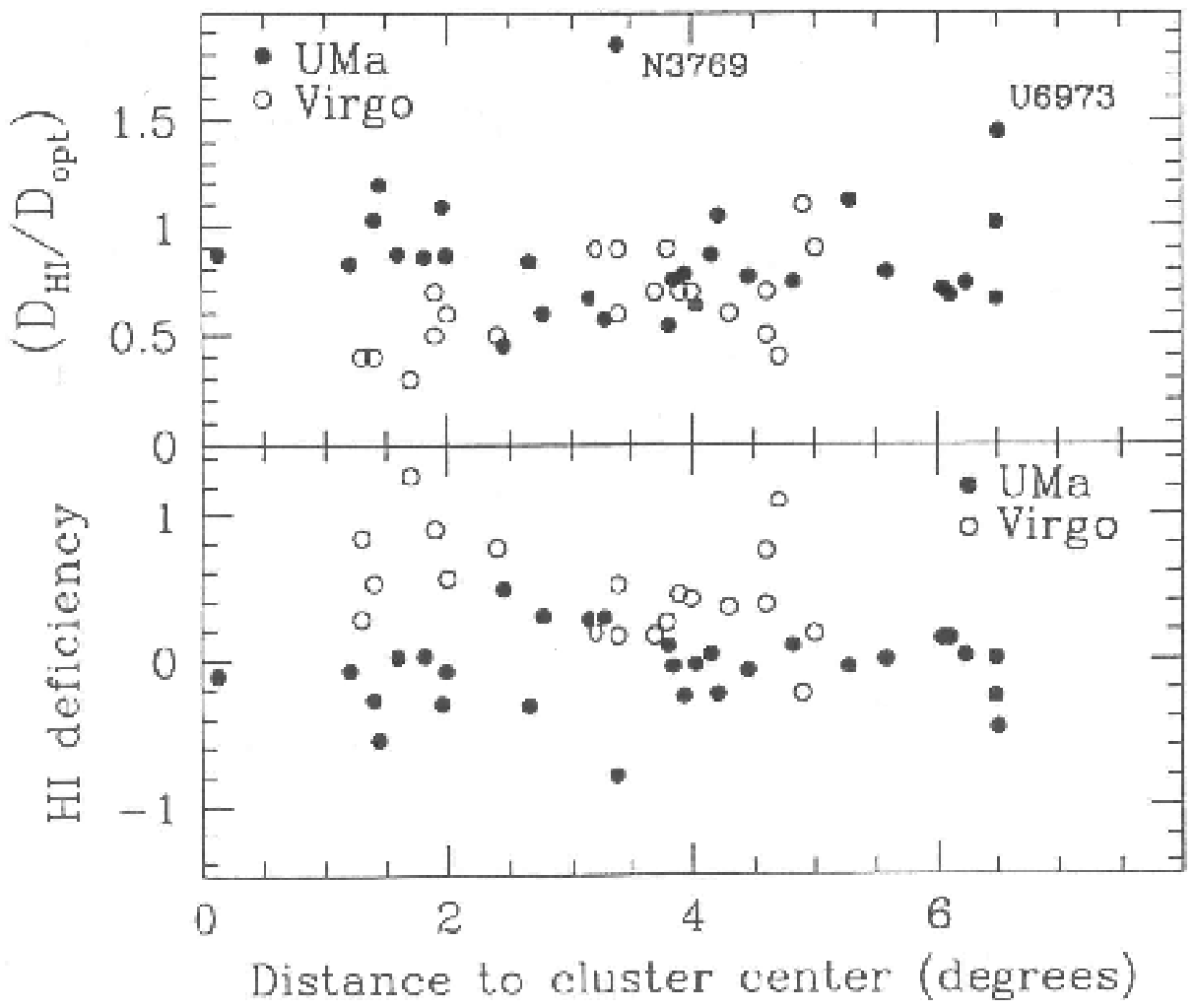}
\end{minipage}

\caption{{\bf Left:} Composition of integrated maps of atomic hydrogen in
the Ursa Major cluster. Individual galaxies are 4$\times$
enlarged. {\bf Right:} A comparison of HI properties of galaxies in the Ursa
Major and Virgo clusters as a function of projected distance to the
cluster center. Above: ratio of HI to optical diameters. Below:
overall HI deficiency as a function projected distance.}

\label{fig:fig1}
\end{figure}

Observing the cold gas in transforming galaxies inherently provides
important clues about the physical processes that lead to its
removal. Many mechanisms have been proposed that could result in the
depletion of the cold gas. Only some are mentioned here.

The fragile extended gas disks of spiral galaxies are effective
tracers of tidal interactions. Tidal interactions between galaxies,
either by a few slow but effective encounters or by many `harassing'
fast encounters (e.g \cite[Moore \etal\ 1996]{MKLDO96}), may result in
the removal of cold gas from the tenuous outer regions, leading to
truncated HI disks. Once an infalling galaxy encounters the hot ICM,
the cold gas may be removed by ram-pressure, and turbulent and viscous
stripping (e.g. \cite[Schulz \& Struck 2001]{SS01}; \cite[Abadi \etal\
1999]{AMB99}), provided the ICM is sufficiently dense and the galaxy
moves fast enough. Ram-pressure may also compress the cold gas clouds,
triggering enhanced star formation. The cold gas could also be removed
by more gentle and slower processes like thermal evaporation of the
ISM by the hot ICM, provided there is sufficient thermal conduction
(e.g \cite[Cowie \& Songaila 1977]{CS77}). The slow removal of the
cold gas, regardless of the physical process, was labeled 'starvation'
by \cite{TEKDSCON03}. Besides the physical removal of the cold gas
from the gravitational potential, a triggered burst of star formation
may quickly consume all the cold gas and lock it up in a new
generation of stars. This exhaustion of the cold gas reservoir can be
achieved within a few$\times$10$^7$ years during a dusty and possibly
obscured star burst. Luckily, the radio continuum emission, a
by-product of HI synthesis imaging, can be used to derive star
formation rates which are unaffected by dust extinction, although
contamination by AGN activity may be significant.

Apart from providing insights in the physical processes mentioned
above, neutral Hydrogen observations may also help to determine the
amount of substructure, the mass accretion rate, and the dynamical
state of galaxy clusters. A volume limited survey reveals the location
of all the gas-rich galaxies, and helps to map the large scale
structure in which these galaxy clusters are embedded. For instance,
if a subclump of galaxies is observed to be gas-rich, it is quite
unlikely it has crossed the high density cluster core on its infall
trajectory. Furthermore, in case the cold gas is in the process of
being removed from the plane of an infalling galaxy, its offset may
indicate the direction in which a galaxy is moving and thus help to
constrain its orbit.

Finally, observed metal abundances of the ICM suggest that at least
some of it finds its origin in the Inter-Stellar Medium (ISM) and it
would be interesting to know how much the effect of ram-pressure
stripping contributes to this.

\section{Galaxy clusters imaged in HI.}\label{sec:sec3}

\begin{figure}

\begin{minipage}{6.5cm}
\includegraphics[height=6.5cm]{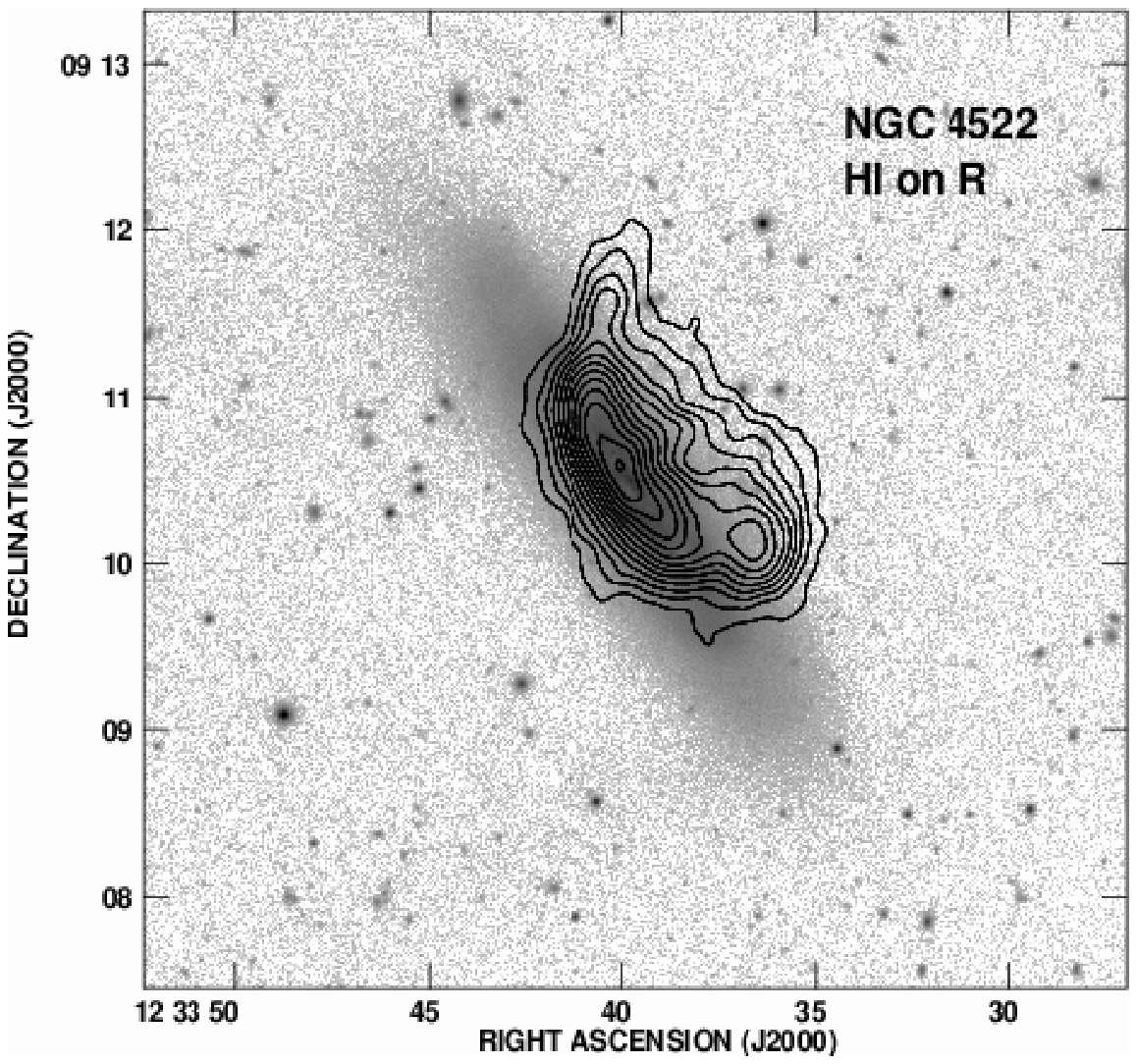}
\end{minipage}

\hspace{8.0cm}
\begin{minipage}{6.5cm}
\vspace{-6.6cm}
\includegraphics[height=6.5cm]{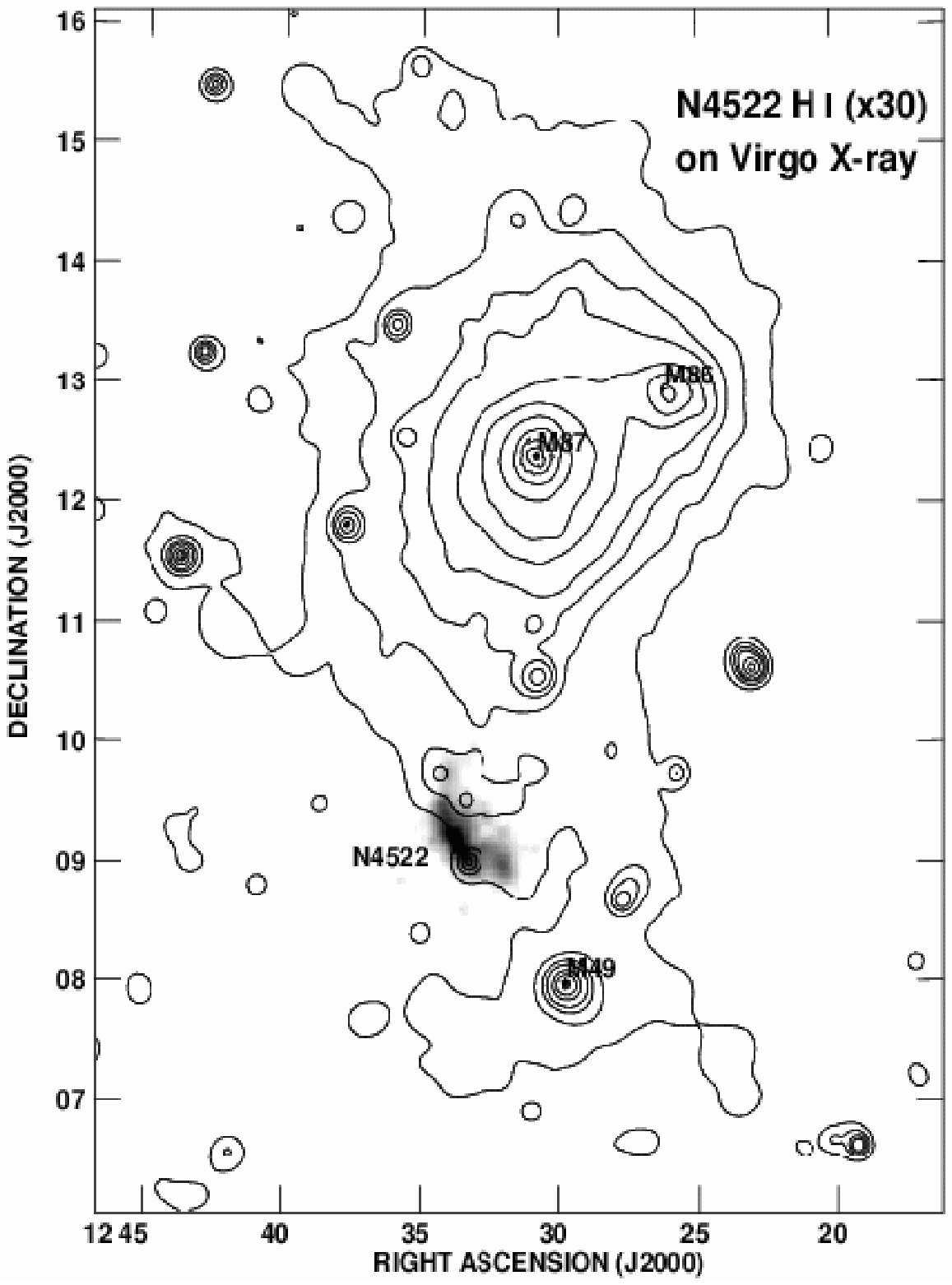}
\end{minipage}

\caption{{\bf Left:} NGC 4522 in the Virgo cluster. The HI contours
are displaced from the optical disk and clearly show the sweeping
effect of ram-pressure stripping. {\bf Right:} The location of NGC
4522 with respect to the distribution of X-ray emitting gas in the
Virgo cluster. Figures taken from Kenney, van Gorkom \& Vollmer (2004).}

\label{fig:fig2}
\end{figure}

In order to investigate the role that the cold gas plays in the
transformation of infalling galaxies, and the influence of the
environment on the various physical processes, a significant sample of
galaxy clusters has been and is being imaged in the 21cm line of
neutral Hydrogen. Table~1 provides an overview of these clusters,
complemented by existing studies from the literature. The clusters
span the full range of redshifts practically accessible by present day
telescopes, up to a redshift of 0.2 where the nearest Butcher-Oemler
clusters are found. The clusters also span a range in richness, X-ray
luminosity and dynamical state. Some clusters display significant
substructure while others, like Abell 963, seem to be fully
relaxed. These clusters and their surroundings encompass a large
variety of environments, ranging from the most dense cluster cores to
areas of significant underdensity. One example of a low density
diffuse cluster is presented in the left panel of Figure~1; Ursa
Major.

\section{Some phenomena observed in HI.}\label{sec:sec4}

Several intriguing and revealing phenomena have been observed in
neutral Hydrogen, and some of these will be illustrated here. First of
all, from spatially unresolved single dish studies, there were strong
hints already in the seventies, that galaxies in cluster cores are
relatively poor in HI compared to their counterparts in the lower
density field. The meaning of this HI deficiency became more clear
with the advent of HI imaging data from the Very Large Array of
galaxies in the Virgo cluster, as illustrated in the right panel of
Figure~1. The gas disks of galaxies within 2 degrees from M87 are
smaller than those of galaxies further away from the cluster
center. At 4 degrees from the cluster center, the gas disks in Virgo
seem to be of comparable size to those in the lower density Ursa Major
environment (upper panel), but the Virgo galaxies are still more
deficient in HI (lower panel); it suggests that the overall column
densities of the gas disks in Virgo are lower compared to the Ursa
Major galaxies. This illustrates that HI imaging gives crucial
physical meaning to the notion of HI deficiency.

\begin{figure}
\includegraphics[width=\textwidth]{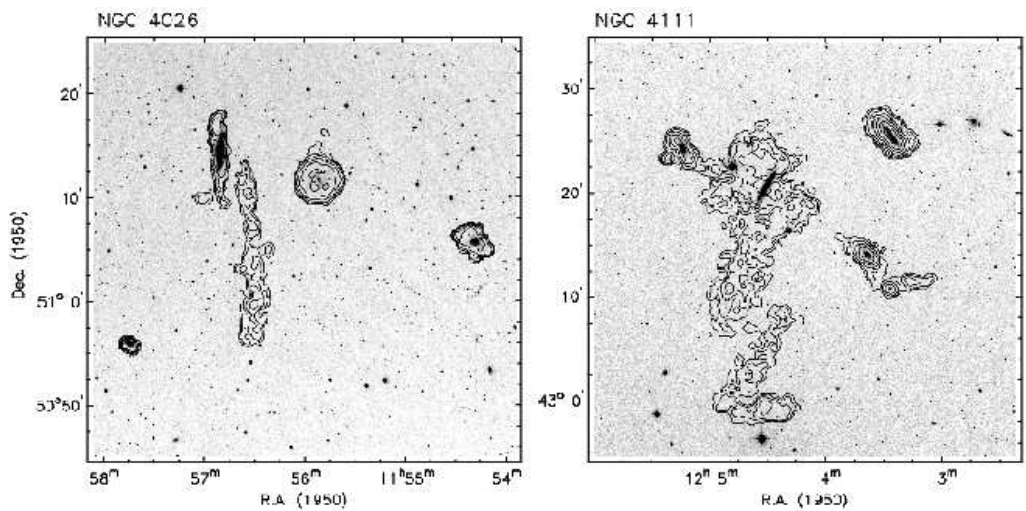}

\caption{HI filaments observed near the brightest lenticular galaxies
in the Ursa Major cluster. These lenticulars are located in small
sub-groups within the Ursa Major volume.}

\label{fig:fig3}
\end{figure}

Early HI imaging data of the Coma and Virgo clusters showed that the
gas disks of galaxies in the outskirts of galaxy clusters can be
asymmetric, both morphologically and kinematically, and significantly
offset from the stellar disk of a galaxy. The spectacular example of
N4522 in the Virgo cluster has been recently published by
\cite{KvGV04} and is reproduced in Figure~2. The small gas disk is
clearly offset from the stellar disk and seems to be bend and swept
away from the plane, interpreted to be the result of ram-pressure
stripping. The right panel of Figure~2 indicates the position of this
galaxy with respect to the X-ray contours. Clearly, the relation
between the location and orbit of this galaxy with respect to the ICM
is not straight forward.

Figure~3 illustrates that ram-pressure stripping is not the only
physical process to remove cold gas from a galaxy. The galaxies N4026
and N4111 are among the brightest lenticulars in the Ursa Major
cluster. This dynamically young cluster has a low velocity dispersion,
lacks a central concentration, shows no X-ray emission from a hot ICM,
contains maybe one elliptical galaxy and a dozen lenticulars. It is an
entirely different environment than the Virgo cluster. The three
brightests lenticular galaxies are located in small subclumps within
the Ursa Major volume, and all three (including N3998) show extended
HI filaments. The orientation and kinematics of these filaments
suggest that they were tidally stripped from the outer disks by their
nearby companions. Do we witness here the transformation of a
late-type spiral galaxy into a lenticular galaxy through tidal
stripping, or the accretion of cold gas from the Inter-Galactic Medium
(IGM)? Detailed modelling is required to answer this
question. However, these lenticular galaxies support the notion that
significant pre-processing of the cold gas component of galaxies may
already occur in the lower density filaments before infalling galaxy
are confronted with the higher density environment of the cluster
outskirts.

Another intriguing observation, not illustrated here, is that the HI
Mass Function (HIMF) may depend significantly on the environment. A
most striking result is that the slope of the HIMF as observed from
the HI Parkes All Sky Survey (HIPASS, \cite[Zwaan \etal\
2003]{Zetal03}) is much steeper ($\sim -1.3$) than what is observed
both in the Local Group (\cite[Zwaan \& Briggs 2000]{ZB00}) and in the
Ursa Major cluster where a flat HIMF slope ($\sim -1.0$) is observed
(\cite[Verheijen \etal\ 2001]{VTTZ01}). Furthermore, the HIPASS survey
suggests there is a morphological dependence of the HIMF, which is not
too surprising.

\begin{figure}
\includegraphics[width=\textwidth]{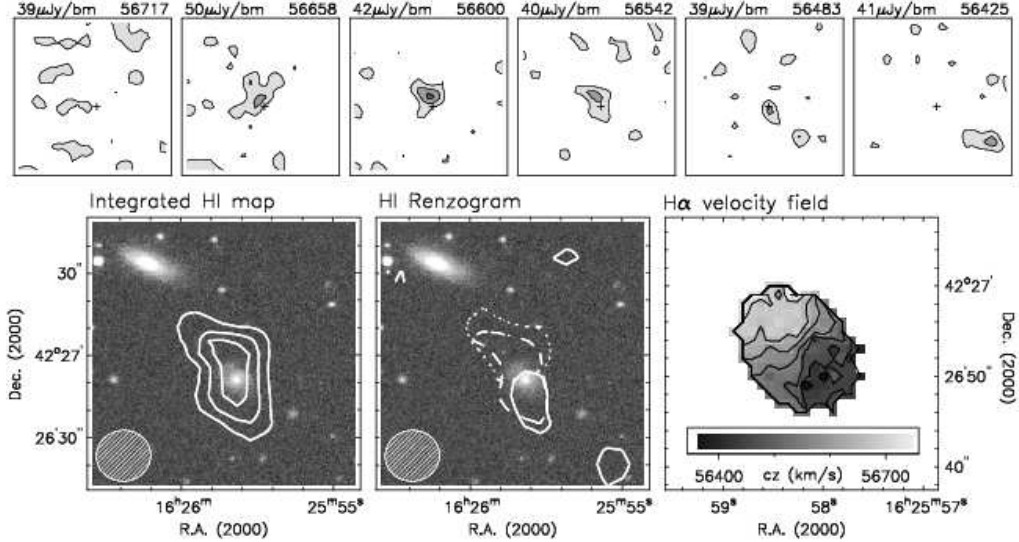}

\caption{Atomic Hydrogen detected in Abell 2192 at z=0.1887. {\bf
Upper panels:} individual channel maps from the VLA datacube. The rms
noise and Heliocentric velocity are noted above above each panel. {\bf
Lower left:} Total HI map. The contours coincide with the position of
an inclined barred late-type galaxy. M$_{\mbox{\scriptsize
HI}}=7\times 10^9$M$_\odot$, SFR$<$12 M$_\odot$/yr. {\bf Lower
middle:} contours from the three channel maps plotted on top of
eachother; solid: cz=56483 km/s, dashed: cz=56542 kms, dotted:
cz=56600 km/s. {\bf Lower left:} H$\alpha$ velocity field of the
optical counterpart obtained with the PMAS IFU spectrograph at the
3.5m telescope on Calar Alto. There is excellent correspondce between
the HI and H$\alpha$ recession velocities, confirming the HI
detection.}

\label{fig:fig4}
\end{figure}

\section{Abell 2192: pushing the HI redshift frontier.}\label{sec:sec5}

In an effort to include a rich galaxy cluster with a clearly observed
Butcher-Oemler effect, several massive clusters near a redshift of 0.2
were selected for HI synthesis observations (see Table~1). This is the
highest redshift accessible, requiring long integration times, and
currently, only pilot studies have been performed to characterize
feasibility and telescope performance. To be able to address issues
related to selection effects, it was decided to observe a less massive
and more diffuse cluster around z=0.2 as well; Abell 2192. Extensive
wide-field optical imaging and multi-object spectroscopy has been
obtained for this cluster. As a feasibility study, A2192 has been
observed with the VLA for $\sim$80 hours, and the resulting data cube
contained at least one HI detection as illustrated in Figure~4. It
concerns a regular inclined barred spiral galaxy at a redshift of
0.1887, located some 15$^\prime$ or a projected 2.8 Mpc from the
cluster center. This galaxy is part of the large scale structure
around A2192. The HI data seem to suggest a possible tidal interaction
with a bright and somewhat lopsided nearby galaxy of yet unknown
redshift, just to the northeast. This highest redshift HI emission
detected to date, demonstrates that HI imaging of Butcher-Oemler
clusters at z=0.2 is feasible with present day facilities.

\section{Conclusions and future plans.}\label{sec:sec6}

Spatially resolved HI synthesis imaging of galaxy clusters provides
crucial insights in the physical processes that trigger and govern the
evolution and transformation of galaxies in the outskirts of
clusters. Spatially resolved data have given physical meaning to HI
deficiencies. The HI morphologies and kinematics of infalling galaxies
help to distinguish ICM-ISM interactions from gravitational
disturbances which are otherwise undetectable. The cold gas reservoir
is a measure of the potential for star formation in a galaxy, and a
close relation is expected between a galaxy's gas content and the
evolutionary state of its stellar population, i.e. starburst, e(a),
post-starburst, or passive. The radio continuum data provide dust-free
measures of star formation rates. Furthermore, volume limited HI
surveys of galaxy clusters help to determine the dynamical state of a
cluster, and illustrates the dynamics of cluster mass accretion.

Feasibility studies for future HI imaging of Butcher-Oemler clusters
at z$\sim$0.2 have yielded the highest redshift HI emission detected
to date at a redshift of 0.1887 in A2192.

Future plans entail the homogenization of the HI data sets for the
various clusters listed in Table~1 in terms of volume and sensitivity.
The continuation of collecting optical wide-field imaging and
spectroscopy necessary to measure the blue galaxy fractions and to
determine the evolutionary states of stellar populations in cluster
galaxies. A new volume limited survey of the Hercules supercluster is
planned in conjunction with wide-field OmegaCam imaging and other
supporting observations.

Finally, the planned Square Kilometer Array (SKA) will allow us to
measure the content, morphology, and kinematics of neutral Hydrogen in
galaxies beyond a redshift of 1. This will revolutionize our
understanding of the physical processes that shape the galaxies and
determine their evolution in every imaginable environment.

\begin{acknowledgments}

The National Radio Astronomy Observatory is a facility of the National
Science Foundation, operated under cooperative agreement by Associated
Universities, Inc.  The Westerbork Synthesis Radio Telescope is
operated by the ASTRON (Netherlands Foundation for Research in
Astronomy) with support from the Netherlands Foundation for Scientific
Research (NWO).  PPak is developed within the ULTROS project under the
German Verbundforschung grant 05AE2BAA/4.  The author wishes to thank
the Scientific Organizing Committee for their invitation and travel
support.

\end{acknowledgments}

%\begin{discussion}

%\end{discussion}

\end{document}